# A comparative study of augmented inverse propensity weighted estimators using outcome-oriented covariate selection via penalization with outcome-adaptive lasso

Wataru Hongo (ORCID: 0000-0002-7439-7704)[1,2*], Shuji Ando (ORCID: 0000-0003-1663-1897)[3], Jun Tsuchida (ORCID: 0000-0003-2349-2709)[4], Takashi Sozu (ORCID: 0000-0002-9857-5819)[5]

[1] Department of Information and Computer Technology, Graduate School of Engineering, Tokyo University of Science, Tokyo, Japan

[2] Advanced Quantitative Sciences, Development, Novartis Pharma K.K., Tokyo, Japan

[3] Department of Information Sciences, Faculty of Science and Technology, Tokyo University of Science, Chiba, Japan

[4] Department of Data Science, Faculty of Data Science, Kyoto Women's University, Kyoto, Japan

[5] Department of Information and Computer Technology, Faculty of Engineering, Tokyo University of Science, Tokyo, Japan

**Abstract**

**Background:**

When estimating causal effects from observational data with numerous covariates, employing penalized covariate selection can improve the estimation efficiency. Outcome-oriented covariate selection, which involves selecting covariates related to the outcome, can enhance efficiency, even for propensity score (PS) methods. For outcome-oriented covariate selection in PS models, outcome-adaptive lasso can be used for penalization with the oracle property. The performance of inverse propensity weighted (IPW) estimators using outcome-adaptive lasso was shown to be superior to that of the IPW estimators using other covariate selection methods for parametric models.

[*] **Corresponding author:** Wataru Hongo (Email: 4622702@ed.tus.ac.jp)

However, the augmented IPW (AIPW) estimator is typically employed as a doubly robust estimator for the average treatment effect, which requires both PS and outcome models. Despite this, which covariate selection method for outcome models should be combined with the outcome-adaptive lasso to form the AIPW estimator remains unclear. We evaluated the performance of the AIPW estimators using the outcome-adaptive lasso for PS models and various outcome-oriented covariate selection via penalization for outcome models.

**Methods:**

We conducted numerical experiments to evaluate the performance of AIPW estimators using various covariate selection via penalization. In numerical experiments, we assessed bias, standard error, and root-mean-squared error. We applied the AIPW estimators to a clinical trial dataset for reference, comparing their point estimates and standard errors obtained via bootstrap.

**Results:**

In the numerical experiments, the performance of the AIPW estimators using outcome-oriented covariate selection via penalization with the oracle property for both PS and outcome models was superior to that of the other estimators and similar to that of the AIPW estimator, which relies on true confounders and outcome predictors. In contrast, the bias of the AIPW estimators not relying on the oracle property was high. In the clinical trial analysis, the AIPW estimators using outcome-oriented covariate selection via penalization with and without the oracle property showed similar estimates and standard errors.

**Conclusions:**

Combining the outcome-adaptive lasso for PS models with outcome-oriented covariate selection via penalization with the oracle property for outcome models produces more unbiased and efficient estimates. No substantial differences in performance were observed among the models.

# Background

Covariate selection is crucial to improve efficiency of causal effect estimation. Ideally, one would determine which covariates to adjust for by constructing a causal directed acyclic graph (DAG) that clearly depicts the causal relationships among variables such as outcome, treatment, and potential covariates [1]. However, in the era of big data, the number of potential covariates can be exceedingly large, making it impractical to capture all possible causal relationships or to apply simple methods such as modified disjunctive cause criterion [2]. This challenge has spurred the development and application of data-driven covariate selection methods. A survey has revealed that more than 10% of observational studies involving variable selection methods and multivariate analysis in five major medical journals employed data-driven variable selection [3]. Nonetheless, the actual methods used are mostly limited to the least absolute shrinkage and selection operator (lasso) [4] and the elastic net [5], largely because these approaches are readily available in existing statistical software packages. Therefore, in this study, we focused on covariate selection methods for parametric models, which can be easily implemented.

Nevertheless, given that the true relationships among the outcome, treatment, and covariates are typically unknown, doubly robust estimators that are consistent if at least one of the propensity score (PS) or outcome models is correctly specified [6–10] are widely recommended [11]. Modern epidemiological studies have gradually adopted such doubly robust methods [12]. The augmented inverse propensity weighted (AIPW) estimator [13–16] is one of the most commonly used doubly robust estimators because it only requires practitioners to do two things they are already comfortable with: (1) specifying a binary regression model for the PS [17] and (2) specifying an outcome model [16]. However, to increase the efficiency of the AIPW estimators, covariate selection must be performed for both PS and outcome models.

Among some covariate selection strategies such as selecting only confounders, outcome-oriented covariate selection, which involves selecting covariates related to the outcome, has been shown to enhance the efficiency of causal effect estimation [18–23], even for PS methods. For outcome-oriented covariate selection in PS models, the outcome-adaptive lasso (OAL) [24] and generalized OAL [25, 26] can be used as penalized logistic regression methods. These methods have the oracle property, similar to the adaptive lasso [27], implying that they select nonzero coefficients with probability tending to one (i.e., sparsity), and nonzero components are estimated as if the true models are known apriori (i.e., asymptotic normality) [28]. For outcome-oriented covariate selection in outcome models, there are numerous penalized covariate selection methods, such as the lasso, elastic net, smoothly clipped absolute deviation (SCAD) [28], adaptive lasso, adaptive elastic net [29], log sum penalty (LSP) [30], and minimax concave penalty (MCP) [31]. In this study, we focused on using the OAL for PS models because the superiority of OAL over other covariate selection methods has been reported in the framework of the inverse propensity weighted (IPW) estimator [24, 32], and it is easy to implement within the adaptive lasso framework. However, it remains unclear which covariate selection method for outcome models should be combined with OAL to form the AIPW estimator.

In this study, we evaluated the performance of the AIPW estimators using OAL for PS models and various outcome-oriented covariate selection via penalization for outcome models, which can be simply referred to as the AIPW estimators using outcome-oriented covariate selection via penalization (hereinafter, AIPW-OCSvP). To this end, we conducted multiple numerical experiments under a range of conditions. In the numerical experiments, we assessed the bias, standard error, and root-mean-squared error to comprehensively evaluate the estimators' performance.

The remainder of this paper is organized as follows. Section 2 (Methods) outlines the foundational methodology, including the definition of the AIPW estimator and how to implement the penalized covariate selection methods that we compare. Section 3 (Numerical experiments) investigates the performance of the methods through a series of numerical experiments, where sample sizes and covariate types are systematically varied. Section 4 (Analysis of a clinical trial example) applies these methods to a clinical trial dataset, illustrating their behavior in a real-world setting. Section 5 (Discussion) concludes the paper by discussing our key findings, the limitations of this paper, and future research directions.

# Methods

### Notation

Let $Y$ be the outcome variable of interest, and let $Z$ be the treatment variable, where $Z = 1$ corresponds to the treatment group and $Z = 0$ corresponds to the control group. Let $Y(1)$ and $Y(0)$ denote the potential outcomes under treatment and control, respectively. Define $X_j$ $(j = 1,2,\dots,p)$ as the $j$-th covariate, and let $\boldsymbol{X}$ be a $p$-dimensional vector of covariates observed prior to $Z$. In the potential outcome framework, we assume the stable unit treatment value assumption (SUTVA) and consistency, implying the outcome $Y$ is whichever potential outcome aligns with the actual treatment (i.e., $Y = ZY(1) + (1 - Z)Y(0)$).

For each participant $i$ $(= 1,2, \ldots, n)$ in an observational study, uppercase letters (e.g., $Y_i, Z_i, \boldsymbol{X}_i$) denote random variables, while the corresponding lowercase letters (e.g., $y_i, z_i, \boldsymbol{x}_i$) denote their realized values. The set of random variables $\{Y_i, Z_i, \boldsymbol{X}_i\}$ is assumed to be independently and identically distributed.

We introduce the partition of a covariate index set based on its relationship with the outcome and treatment:

1. $\mathcal{C}$ (confounders): variables related to both the outcome and the treatment.
2. $\mathcal{P}$ (outcome predictors): variables related only to the outcome.
3. $\mathcal{I}$ (instrumental variables): variables related only to the treatment.
4. $\mathcal{S}$ (spurious variables): variables related to neither the outcome nor the treatment.

The outcome-oriented covariate selection targets the variables related to the outcome, namely $j \in \mathcal{C} \cup \mathcal{P}$.

**Estimation of the average treatment effect**

Treatment effects are defined based on the contrast between $Y(1)$ and $Y(0)$. We focus on the popular measure of treatment effect [33], known as the average treatment effect (ATE), defined as $\theta = E[Y(1)] - E[Y(0)]$ [17]. In observational studies, four assumptions are generally required to ensure unbiased estimation of the ATE: SUTVA, consistency, unmeasured confounding (or ignorability), and positivity (i.e., $P(Z = z|\boldsymbol{X} = \boldsymbol{x}) > 0$ for $Z = 0,1$). Unmeasured confounding, typically written as $Z \perp Y(z)|\boldsymbol{X} = \boldsymbol{x}$, is pivotal for identifying causal effects from observational data. A sufficient set of covariates to satisfy unmeasured confounding is given by $j \in \mathcal{C}$.

We first illustrate the IPW estimator [18] as an example of an ATE estimator based on PS. Let $\boldsymbol{\alpha}$ be the parameter of the PS model and let $\hat{\boldsymbol{\alpha}}$ be its estimator. For participant $i$, define $e(\boldsymbol{x}_i; \hat{\boldsymbol{\alpha}})$ as the estimated PS. The IPW estimator $\hat{\theta}^{IPW}$ is then given by

$$\hat{\theta}^{IPW} = \frac{1}{\sum_{i=1}^{n}\left(\frac{z_i}{e(\boldsymbol{x}_i;\widehat{\boldsymbol{\alpha}})}\right)}\sum_{i=1}^{n}\left(\frac{z_i y_i}{e(\boldsymbol{x}_i;\widehat{\boldsymbol{\alpha}})}\right) - \frac{1}{\sum_{i=1}^{n}\left(\frac{1-z_i}{1-e(\boldsymbol{x}_i;\widehat{\boldsymbol{\alpha}})}\right)}\sum_{i=1}^{n}\left(\frac{(1-z_i)y_i}{1-e(\boldsymbol{x}_i;\widehat{\boldsymbol{\alpha}})}\right).$$

The IPW estimator is consistent for the ATE, if the PS model is correctly specified.

Subsequently, we describe the g-computation estimator as an example of an ATE estimator based on an outcome regression. Let $\boldsymbol{\beta}_{Z=z}$ be the parameter vector for the outcome regression model $Y = \boldsymbol{X}\boldsymbol{\beta}_{Z=z} + \epsilon$, where $\epsilon$ is the error term, and let $\widehat{\boldsymbol{\beta}}_{Z=z}$ be its estimator. We estimate $\boldsymbol{\beta}_{Z=z}$ separately using data from participants in each treatment group $Z = z$. For participant $i$, define $m(\boldsymbol{x}_i;\widehat{\boldsymbol{\beta}}_{Z=z})$ as the predicted value of $Y$ under $Z = z$. Then, the g-computation estimator of ATE, $\hat{\theta}^{gComp}$, is given by

$$\hat{\theta}^{gComp} = \frac{1}{n}\sum_{i=1}^{n} m(\boldsymbol{x}_i;\widehat{\boldsymbol{\beta}}_{Z=1}) - \frac{1}{n}\sum_{i=1}^{n} m(\boldsymbol{x}_i;\widehat{\boldsymbol{\beta}}_{Z=0}).$$

The g-computation estimator is consistent for the ATE, if the outcome models are correctly specified.

The drawback of relying solely on either the PS model or the outcome models is that the resulting estimator is inconsistent when that model is mis-specified (e.g., incorrect functional form or omission of confounders in $C$). To address this issue, the AIPW estimator [16] can be used. The AIPW estimator $\hat{\theta}^{AIPW}$ is given by

$$\hat{\theta}^{AIPW} = \frac{1}{n}\sum_{i=1}^{n}\left[\frac{z_i y_i}{e(\boldsymbol{x}_i;\widehat{\boldsymbol{\alpha}})} - \frac{z_i - e(\boldsymbol{x}_i;\widehat{\boldsymbol{\alpha}})}{e(\boldsymbol{x}_i;\widehat{\boldsymbol{\alpha}})} m(\boldsymbol{x}_i;\widehat{\boldsymbol{\beta}}_{Z=1})\right]$$
$$- \frac{1}{n}\sum_{i=1}^{n}\left[\frac{(1-z_i)y_i}{1-e(\boldsymbol{x}_i;\widehat{\boldsymbol{\alpha}})} + \frac{z_i - e(\boldsymbol{x}_i;\widehat{\boldsymbol{\alpha}})}{1-e(\boldsymbol{x}_i;\widehat{\boldsymbol{\alpha}})} m(\boldsymbol{x}_i;\widehat{\boldsymbol{\beta}}_{Z=0})\right].$$

This estimator is consistent if at least one of the PS or outcome models is correctly specified, and it achieves smaller asymptotic variance than the IPW estimator, when both models are correctly specified [14,18].

**Penalized covariate selection methods**

We used lasso and OAL for parameter estimation in the PS model assuming a logistic regression. Additionally, we used lasso, elastic net, SCAD, adaptive lasso, adaptive elastic net, LSP, and MCP for parameter estimation in the outcome models assuming a linear regression. Table 1 summarizes the ATE estimators used in the numerical experiments, along with the methods used to estimate the parameters in the PS and outcome models. The AIPW estimators using the OAL for the PS model are classified as AIPW-OCSvP. We used IPW-OAL and gComp-AdL as reference values, each relying solely on either the PS model or the outcome models. We used AIPW-Targ, which relies on the outcome-oriented covariates $j \in \mathcal{C} \cup \mathcal{P}$, as a target of AIPW-OCSvP.

Under the logistic model for the PS, the OAL estimates $\widehat{\boldsymbol{\alpha}}(OAL)$ are obtained by solving

$$\widehat{\boldsymbol{\alpha}}(OAL) = \underset{\boldsymbol{\alpha}}{\operatorname{argmin}} \left[ \sum_{i=1}^{n} \{-z_i(\boldsymbol{x}_i^T \boldsymbol{\alpha}) + \log(1 + \exp(\boldsymbol{x}_i^T \boldsymbol{\alpha}))\} + \lambda \sum_{j=1}^{d} \widehat{\omega}_j |\alpha_j| \right],$$

where $\widehat{\omega}_j = |\tilde{\beta}_j|^{-\gamma}$ with $\gamma > 1$, and $\tilde{\beta}_j$ is the ordinary least squares estimate for $X_j$ obtained from a linear regression of $Y$ on $Z$ and $\boldsymbol{X}$. OAL is a straightforward variant of the ordinary adaptive lasso that incorporates information from the outcome model when computing the weights and thus can be implemented using existing statistical packages such as the *glmnet* package in R.

The lasso was implemented using the cv.glmnet function of *glmnet* package in R with default arguments. OAL was implemented using the cv.glmnet function to select hyperparameters, following the method outlined by Shortreed and Ertefaie [24]. Adaptive lasso (AdL) for outcome models was implemented by the cv.glment function with the penalty factors of $|\widehat{\boldsymbol{\beta}}(\text{ols})|^{-1}$, where $\widehat{\boldsymbol{\beta}}(\text{ols})$ was the ordinary least squares estimate obtained from the same outcome regression model. Elastic net (EN) was implemented by the cv.glment function using the elastic net mixing parameter $\alpha = 0.5$. Notably, the value of $\alpha$ of 0 or 1 corresponds to the ridge or lasso penalty, respectively. Adaptive elastic net (AEN) was implemented on the aenet function of *msaenet* package in R using the elastic net mixing parameter $\alpha = \{0.2, 0.4, \ldots, 0.8\}$. SCAD and MCP were implemented using the cv.ncvreg function of the *ncvreg* package in R, with default arguments. LSP was implemented by iterating adaptive lasso updates: we started with an initial weights $\widehat{w}_j^{(0)} = 1$ for all $j$ and updated the weights $\widehat{\boldsymbol{w}}^{(l+1)} = 1/(|\widehat{\boldsymbol{\beta}}^{(l)}| + 0.0001)$, repeating up to $l_{\max} = 4$ [30]. We selected the hyperparameters for each penalized covariate selection method by 10-fold cross-validation with default loss functions, except in the case of OAL.

**Table 1** ATE estimators used in the numerical experiments

| Estimator name | ATE estimator | Variable selection method for PS model | Variable selection method for outcome model |
|---|---|---|---|
| naive | Difference in means of $Y$ | - | - |
| IPW-OAL | IPW | OAL | - |
| gComp-AdL | g-computation | - | adaptive lasso |
| AIPW-Targ | AIPW | Ordinary least squares using $x_j$ ($j \in \mathcal{C} \cup \mathcal{P}$) | Ordinary least squares using $x_j$ ($j \in \mathcal{C} \cup \mathcal{P}$) |
| AIPW-Las-Las | AIPW | lasso | lasso |
| AIPW-OAL-Las | AIPW | OAL | lasso |
| AIPW-OAL-EN | AIPW | OAL | elastic net |
| AIPW-OAL-AdL | AIPW | OAL | adaptive lasso |
| AIPW-OAL-AEN | AIPW | OAL | adaptive elastic net |
| AIPW-OAL-SCAD | AIPW | OAL | SCAD |
| AIPW-OAL-LSP | AIPW | OAL | log sum penalty |
| AIPW-OAL-MCP | AIPW | OAL | MCP |
| AIPW-Farrell* | AIPW | lasso | lasso |

* The estimator proposed by Farrell [34] uses lasso only to select covariates to be adjusted for and then fits unpenalized PS and outcome models with the selected covariates for the AIPW estimator.

# Numerical experiments

### Setup

Suppose that each participant $i$ has a covariate vector $\boldsymbol{X}_i$, which is independently and identically distributed from a $p$-dimensional multivariate normal distribution with mean vector $\boldsymbol{\mu} = \boldsymbol{0}$, every variance equal to 1, and every correlation coefficient $\rho$. The binary treatment variable $Z_i$ and the continuous outcome variable $Y_i$ are generated as follows:

$$\text{logit}\{\Pr(Z_i = 1|\boldsymbol{X}_i)\} = \sum_{j=1}^{p} \alpha_j X_{ij}, Y_i = 0.5Z_i + \sum_{j=1}^{p} \beta_j X_{ij} + \epsilon_i,$$

where $\epsilon_i$ is an error term denoting an independently and identically distributed random variable from a standard normal distribution. To evaluate the performance under strong and weak confounding, we specify $\boldsymbol{\alpha}$ and $\boldsymbol{\beta}$ as follows:

Strong confounding: $\boldsymbol{\alpha} = \{1,1,0,0,1,1,0,\dots,0\}, \boldsymbol{\beta} = \{0.6,0.6,0.6,0.6,0,0,0,\dots,0\}$,

Weak confounding: $\boldsymbol{\alpha} = \{0.4,0.4,0,0,1,1,0,\dots,0\}, \boldsymbol{\beta} = \{0.2,0.2,0.6,0.6,0,0,0,\dots,0\}$.

Under weak confounding, identifying the true confounders is more difficult than under strong confounding. In all scenarios, $\mathcal{C} = \{1,2\}$, $\mathcal{P} = \{3,4\}$, $\mathcal{I} = \{5,6\}$, and $\mathcal{S} = \{7,\dots,p\}$.

To examine model mis-specification, we introduce the transformed variables $\boldsymbol{U} = \{3X_1/(1 + \exp X_2), 5\sin X_2, X_3^3/3, 5\sin X_4, (X_5 + X_6)/\sqrt{2}, 5\sin X_6, X_7, \dots, X_p\}$. As an example, we used $\boldsymbol{U}$ to generate the treatment variable instead of $\boldsymbol{X}$ and $\boldsymbol{X}$ to generate the outcome variable in the scenario where the PS model was mis-specified but the outcome models were correctly specified. The transformed variable $\boldsymbol{U}$ is unobservable; therefore, the PS and outcome regression models always include $\boldsymbol{X}$ as exploratory variables, even when composing the AIPW-Targ estimator.

To assess performance under binary covariates, we used the transformed covariates $\boldsymbol{X}_{binary}$ to evaluate the performance under binary covariates. $X_{binary,j}$ is set to $1$ if $X_j > 0$; otherwise, $X_{binary,j}$ is set to $-1$. All covariates are standardized via Z-score normalization prior to performing penalized regression.

We consider $(n, p) \in \{(200,80), (500,200), (1000, 400)\}$ for the sample size $(n)$ and number of covariates $(p)$. We investigated three correlation levels $\rho \in \{0,0.2,0.5\}$ in the scenarios where both PS and outcome models can be correctly specified while we used one correlation level $\rho = 0$ in the other scenarios. Each condition was replicated 1000 times.

# Results

We fixed $\rho = 0$ and only presented the results for $n = 200$ under strong confounding and with continuous and binary covariates owing to space constraints. We describe all results of the numerical experiments (72 different settings) in Additional File 1. Additional File 1 is available on

https://github.com/WataruHongo/SupplMat.

Figures 1 and 2 graphically present the ATE estimates of each method for 1000 simulations with continuous and binary covariates, respectively. The dashed line indicates the true ATE of 0.5. The white circles represent the mean values of 1000 ATE estimates. The naive estimates are excluded from the figures because they significantly differed from the other estimates. The boxes for AIPW-OCSvP with the oracle property for the PS and outcome models are highlighted in gray in Figures 1 and 2. Table 2 presents the bias, standard error, and root-mean-squared error for the ATE estimators with (A) continuous and (B) binary covariates. Values exceeding $10^8$ or falling below $-10^8$ are displayed with a hyphen in Table 2.

Across different types of model specifications (scenarios a, b, c, and d) and covariates (continuous and binary), the bias of AIPW-OCSvP with the oracle property (AIPW-OAL-AdL, AIPW-OAL-AEN, AIPW-OAL-SCAD, AIPW-LSP, and AIPW-OAL-MCP) was smaller than that of AIPW-OCSvP without the oracle property (AIPW-OAL-Las and AIPW-OAL-EN). This trend held true even when increasing the sample size and number of covariates or when increasing the correlation among covariates, regardless of the strength of confounding (see Table A.1–A.12 in Additional File 1). Under strong confounding, SCAD and MCP tended not to include the variables unrelated to the outcome more effectively than the other methods with the oracle property (AdL, AEN, and LSP). Consequently, the bias and standard errors of AIPW-OAL-SCAD and AIPW-OAL-MCP were slightly smaller than those of the other AIPW-OCSvP with the oracle property (AIPW-OAL-AdL, AIPW-OAL-AEN, and AIPW-OAL-LSP). In contrast, under weak confounding, true confounders were less likely to be selected in SCAD and MCP, leading to slightly larger bias in AIPW-OAL-SCAD and AIPW-OAL-MCP than in other AIPW estimators (see Table A.7–A.12 in Additional File 1). Overall, when constructing AIPW estimators by combining the OAL with any of OCSvP with the oracle property for the outcome models (AdL, AEN, SCAD, LSP, or MCP), the differences in performance among these estimators were not substantial.

The performance of AIPW-OCSvP with the oracle property was the closest to AIPW-Targ, which exclusively relies on outcome-oriented covariates. Simply using covariate selection via penalization without the oracle property (AIPW-Las-Las, AIPW-OAL-Las, and AIPW-OAL-EN) introduced bias owing to the shrinkage of regression coefficients. In particular, the bias of AIPW-Las-Las was large across all scenarios (a, b, c, and d), while the bias of AIPW-OAL-Las and AIPW-OAL-EN was relatively large under scenarios where the outcome model was mis-specified (scenario c). The performance gap between AIPW estimators with and without the oracle property was the most noticeable for small sample sizes but persisted even when $n = 1000$ (see Table A.5, A.6, A.11, and A.12 in Additional File 1). Because AIPW-Farrell adjusts for the same set of covariates as AIPW-Las-Las, its PS model includes variables strongly associated with treatment without shrinkage bias owing to its refitting approach. However, this covariate selection strategy for AIPW-Farrell leads to larger standard errors compared with AIPW-OCSvP. Moreover, unstable estimates of AIPW-Farrell were occasionally observed, yielding very large or very small estimates.

In most scenarios, the performance of AIPW-OCSvP with the oracle property was superior compared to that of the other estimators in terms of both bias and standard errors. However, under strong confounding scenarios with continuous covariates, where both the PS and outcome models were mis-specified, the bias of AIPW-OCSvP with the oracle property sometimes tended to be larger than that of IPW-OAL or gComp-AdL. Nonetheless, the advantage of IPW-OAL or gComp-AdL in these situations was typically modest. Overall, the findings from the numerical experiments in this study highlight that AIPW-OCSvP with the oracle property is robust and effective across a wide range of conditions.

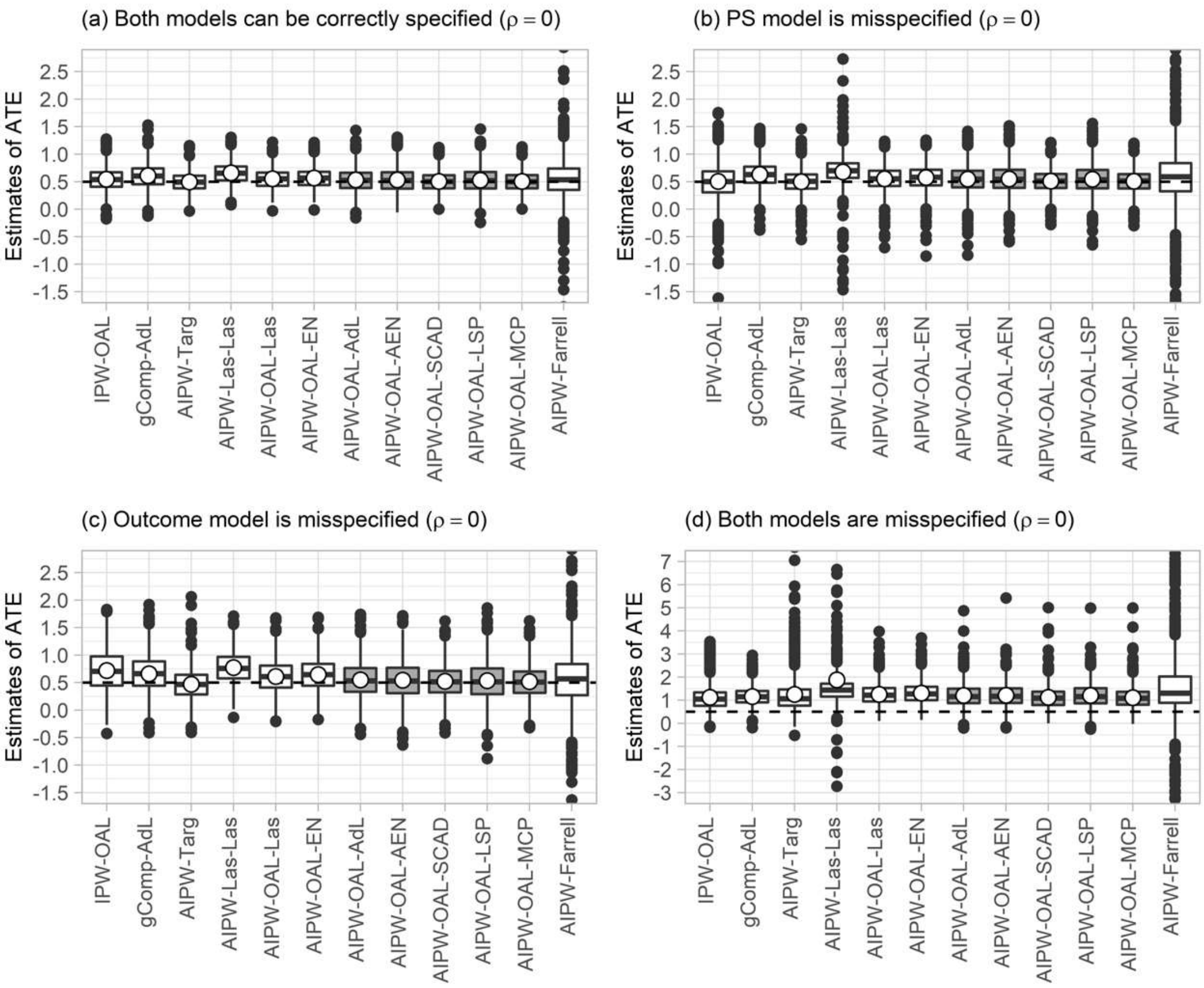

**Fig. 1** Box plots of 1000 estimates for the average treatment effect (ATE) with continuous covariates

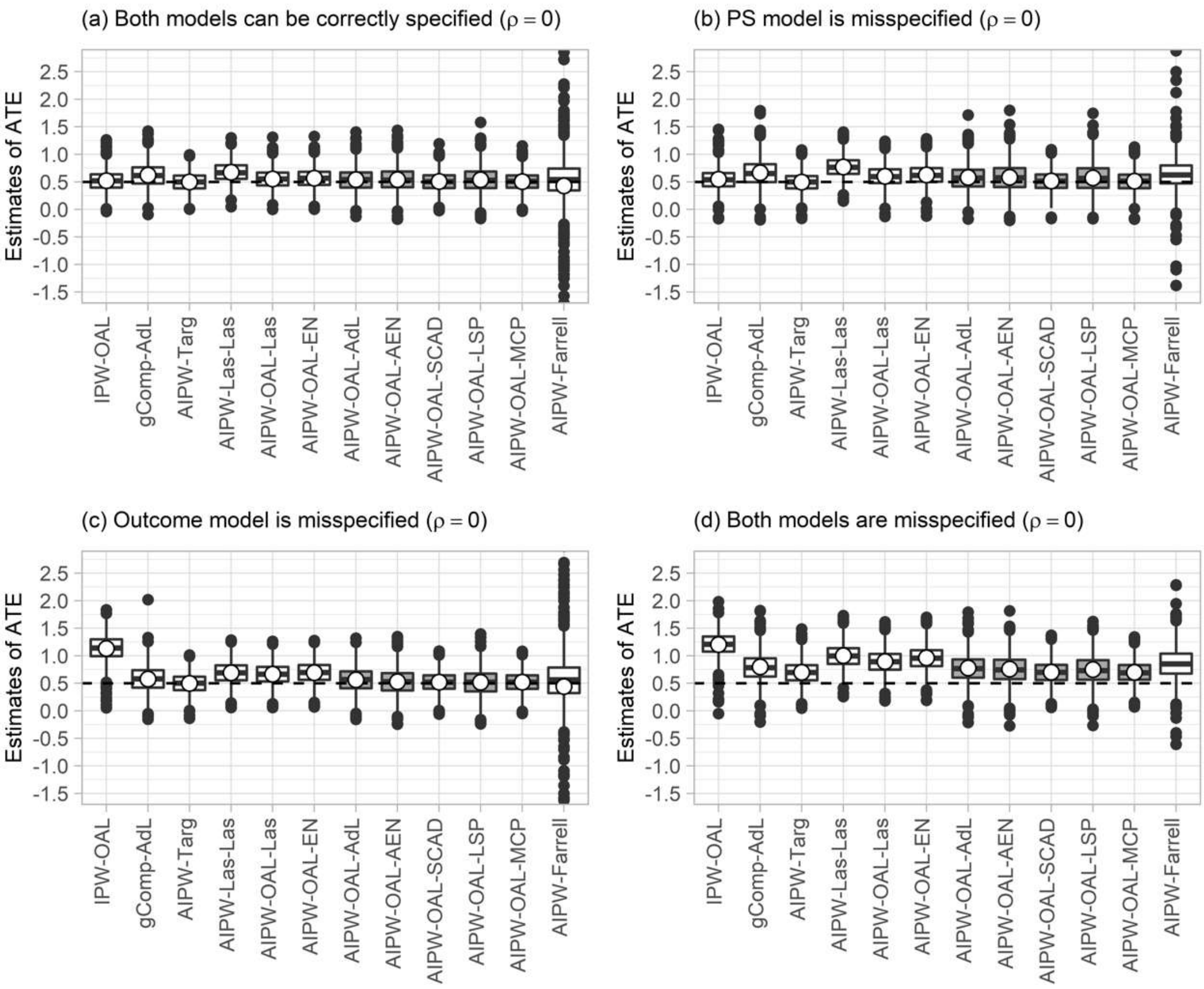

**Fig. 2** Box plots of 1000 estimates for the average treatment effect (ATE) with binary covariates

**Table 2** Results of numerical experiments

| | (a) | | | (b) | | | (c) | | | (d) | | |
|---|---|---|---|---|---|---|---|---|---|---|---|---|
| | Bias | SE | RMSE | Bias | SE | RMSE | Bias | SE | RMSE | Bias | SE | RMSE |
| *(A) Continuous covariates* | | | | | | | | | | | | |
| naive | 0.730 | 0.223 | 0.763 | 0.731 | 0.216 | 0.762 | 1.687 | 0.451 | 1.746 | 2.179 | 0.437 | 2.223 |
| IPW-OAL | 0.045 | 0.203 | 0.208 | 0.005 | 0.356 | 0.356 | 0.220 | 0.384 | 0.442 | 0.623 | 0.596 | 0.863 |
| gComp-AdL | 0.109 | 0.220 | 0.245 | 0.131 | 0.236 | 0.271 | 0.160 | 0.332 | 0.368 | 0.658 | 0.401 | 0.770 |
| AIPW-Targ | -0.005 | 0.173 | 0.173 | -0.002 | 0.215 | 0.215 | -0.030 | 0.321 | 0.322 | 0.749 | 0.843 | 1.128 |
| AIPW-Las-Las | 0.161 | 0.186 | 0.246 | 0.178 | 0.801 | 0.821 | 0.271 | 0.285 | 0.394 | 1.377 | 9.852 | 9.948 |
| AIPW-OAL-Las | 0.054 | 0.184 | 0.192 | 0.054 | 0.230 | 0.236 | 0.112 | 0.296 | 0.317 | 0.753 | 0.448 | 0.876 |
| AIPW-OAL-EN | 0.067 | 0.186 | 0.197 | 0.074 | 0.235 | 0.247 | 0.145 | 0.298 | 0.331 | 0.808 | 0.440 | 0.920 |
| AIPW-OAL-AdL | 0.033 | 0.211 | 0.213 | 0.046 | 0.256 | 0.260 | 0.049 | 0.330 | 0.333 | 0.704 | 0.498 | 0.862 |
| AIPW-OAL-AEN | 0.031 | 0.223 | 0.225 | 0.049 | 0.266 | 0.270 | 0.042 | 0.350 | 0.352 | 0.713 | 0.514 | 0.879 |
| AIPW-OAL-SCAD | 0.006 | 0.180 | 0.180 | 0.009 | 0.200 | 0.201 | 0.023 | 0.289 | 0.290 | 0.620 | 0.476 | 0.782 |
| AIPW-OAL-LSP | 0.026 | 0.225 | 0.227 | 0.043 | 0.273 | 0.276 | 0.033 | 0.353 | 0.355 | 0.701 | 0.524 | 0.875 |
| AIPW-OAL-MCP | 0.005 | 0.178 | 0.178 | 0.007 | 0.203 | 0.203 | 0.016 | 0.286 | 0.286 | 0.613 | 0.472 | 0.774 |
| AIPW-Farrell | - | - | - | - | - | - | - | - | - | - | - | - |
| *(B) Binary covariates* | | | | | | | | | | | | |
| naive | 0.754 | 0.215 | 0.784 | 0.793 | 0.208 | 0.820 | 2.156 | 0.518 | 2.217 | 2.576 | 0.521 | 2.628 |
| IPW-OAL | 0.023 | 0.192 | 0.193 | 0.047 | 0.201 | 0.207 | 0.633 | 0.238 | 0.676 | 0.703 | 0.223 | 0.738 |
| gComp-AdL | 0.123 | 0.220 | 0.252 | 0.165 | 0.256 | 0.305 | 0.078 | 0.227 | 0.240 | 0.297 | 0.263 | 0.396 |
| AIPW-Targ | -0.002 | 0.166 | 0.166 | -0.004 | 0.167 | 0.167 | -0.002 | 0.193 | 0.193 | 0.199 | 0.201 | 0.283 |
| AIPW-Las-Las | 0.174 | 0.182 | 0.252 | 0.271 | 0.189 | 0.330 | 0.190 | 0.204 | 0.278 | 0.501 | 0.220 | 0.547 |
| AIPW-OAL-Las | 0.052 | 0.182 | 0.190 | 0.102 | 0.194 | 0.219 | 0.165 | 0.194 | 0.255 | 0.395 | 0.214 | 0.449 |
| AIPW-OAL-EN | 0.067 | 0.183 | 0.195 | 0.127 | 0.195 | 0.233 | 0.193 | 0.199 | 0.277 | 0.463 | 0.222 | 0.513 |
| AIPW-OAL-AdL | 0.041 | 0.215 | 0.219 | 0.080 | 0.246 | 0.259 | 0.063 | 0.222 | 0.230 | 0.278 | 0.262 | 0.382 |
| AIPW-OAL-AEN | 0.043 | 0.230 | 0.234 | 0.084 | 0.270 | 0.283 | 0.028 | 0.236 | 0.237 | 0.257 | 0.280 | 0.380 |
| AIPW-OAL-SCAD | 0.006 | 0.179 | 0.179 | 0.015 | 0.184 | 0.185 | 0.022 | 0.193 | 0.194 | 0.203 | 0.207 | 0.289 |
| AIPW-OAL-LSP | 0.039 | 0.236 | 0.239 | 0.072 | 0.276 | 0.285 | 0.018 | 0.236 | 0.236 | 0.250 | 0.280 | 0.375 |
| AIPW-OAL-MCP | 0.006 | 0.179 | 0.179 | 0.013 | 0.182 | 0.183 | 0.020 | 0.193 | 0.194 | 0.199 | 0.205 | 0.286 |
| AIPW-Farrell | -0.072 | 2.957 | 2.958 | - | - | - | -0.058 | 2.368 | 2.369 | - | - | - |

Bias, standard error (SE), and root-mean-squared error (RMSE) for the ATE estimators with (A) continuous covariates and (B) binary covariates under the strong confounding scenarios where (a) PS and outcome models can be correctly specified, (b) PS model is mis-specified, (c) outcome model is mis-specified, and (d) both PS and outcome models are mis-specified. Values exceeding $10^8$ or falling below $-10^8$ are displayed with a hyphen.

# Analysis of a clinical trial example

We applied the methods shown in Table 1 (except AIPW-Targ) to the right heart catheterization (RHC) dataset from a clinical trial investigating the effect of RHC on the length of the intensive care unit (ICU) stay [35]. We obtained the RHC dataset from https://hbiostat.org/data/, provided by the Vanderbilt University Department of Biostatistics. It originally included 5735 critically ill patients who received or did not receive RHC in the ICU. The sample size analyzed was 5734, excluding one case in which the outcome variable of the number of days in the ICU was missing. A total of 2183 patients received RHC ($Z = 1$), while 3551 patients did not undergo RHC ($Z = 0$). We used 68 variables as covariates, including 21 continuous variables and 47 binary variables (including dummy variables), excluding four binary variables where fewer than 0.5 % of patients fell into one category.

Table 3 shows ATE estimates of the effect of RHC on the log-transformed number of days in the ICU and their 95% confidence intervals by generating 1000 bootstrap samples and then determining the 2.5$^{th}$ and 97.5$^{th}$ percentiles of the resulting distribution.

The estimate of the IPW-OAL was closer to the naive estimate and relatively higher than that of gComp-AdL and AIPW estimators. All AIPW estimators produced nearly identical estimates and standard errors, although the standard errors of AIPW-Las-Las and AIPW-Farrell were slightly higher than those of the other AIPW estimators. This may be because the PS model for these methods tends to include variables that are strongly related to the treatment.

**Table 3** Results of the RHC dataset analysis

| Estimator | ATE | Standard error | 95% CI |
|---|---|---|---|
| naive | 0.213 | 0.027 | (0.162–0.263) |
| IPW-OAL | 0.148 | 0.029 | (0.084–0.200) |
| gComp-AdL | 0.119 | 0.029 | (0.060–0.171) |
| AIPW-Las-Las | 0.126 | 0.033 | (0.064–0.184) |
| AIPW-OAL-Las | 0.126 | 0.031 | (0.060–0.182) |
| AIPW-OAL-EN | 0.126 | 0.031 | (0.060–0.181) |
| AIPW-OAL-AdL | 0.125 | 0.031 | (0.062–0.182) |
| AIPW-OAL-AEN | 0.124 | 0.032 | (0.058–0.184) |
| AIPW-OAL-SCAD | 0.121 | 0.032 | (0.057–0.184) |
| AIPW-OAL-LSP | 0.119 | 0.032 | (0.058–0.184) |
| AIPW-OAL-MCP | 0.128 | 0.032 | (0.058–0.185) |
| AIPW-Farrell | 0.130 | 0.036 | (0.063–0.194) |

# Discussion

In this study, we evaluated the performance of AIPW-OCSvP. Our numerical experiments showed that the performance of AIPW-OCSvP with the oracle property for both the PS and outcome models such as the OAL for the PS model and adaptive lasso for the outcome models was superior compared to that of the other estimators in terms of bias and variance, even under small sample sizes. These findings are consistent with those of Witte and Didelez [22], who found that doubly robust estimators with outcome-oriented covariate selection perform well. In contrast, the bias of AIPW estimators without the oracle property (particularly applying lasso to both the PS and outcome models) was significantly large, likely because the oracle property effectively mitigates the shrinkage bias inherent in methods like the lasso or EN.

Among OCSvP with the oracle property for the outcome models, we recognize that non-convex penalty functions such as SCAD and MCP may pose challenges in practice. These approaches can have multiple local optima, making it challenging to select parameters to achieve the oracle property [36, 37]. Nevertheless, in our experiments, the performance gap between the AIPW estimators using SCAD or MCP and that using the other OCSvP was not substantial.

Although our study also focused on AIPW-OCSvP largely owing to the ease of implementation, other doubly robust estimators with variable selection exist. The doubly robust matching estimator with lasso [38] seems less sensitive to PS and outcome model mis-specification, potentially addressing scenarios where the bias of AIPW estimators is larger than that of the IPW or g-computation estimator. GLiDeR [39]—an AIPW estimator using a variant of group lasso—may serve as a direct comparator to our approach because it emphasizes outcome-oriented covariate selection. Furthermore, the targeted maximum likelihood estimator [40] combined with advanced machine learning and without explicit covariate selection might yield superior performance in some settings.

In summary, our results show that the performance of AIPW-OCSvP with the oracle property for both the PS and outcome models is superior to that of the other AIPW estimators across a wide range of conditions. Our numerical experiments indicated that switching among different OCSvP with the oracle property for the outcome model did not substantially affect performance as long as using the OAL for the PS model, suggesting that practitioners can generally expect robust results from any OCSvP with the oracle property for the outcome model, provided that they are comfortable with its implementation. However, while AIPW-OCSvP with the oracle property is straightforward to implement, it remains unclear whether the performance surpasses that of other doubly robust estimators. Therefore, comparing these estimators on an even broader array of doubly robust methods under various conditions will be an important direction for future research.

# List of abbreviations

AdL, adaptive lasso; AEN, adaptive elastic net; AIPW, augmented inverse propensity weighted; ATE, average treatment effect; DAG, directed acyclic graph; EN, elastic net; ICU, intensive care unit; IPW, inverse propensity weighted; LSP, log sum penalty; MCP, minimax concave penalty; OAL, outcome-adaptive lasso; PS, propensity score; RHS, right heart catheterization; SCAD, smoothly clipped absolute deviation; SUTVA, stable unit treatment value assumption

# Declarations

### Ethics approval and consent to participate

Not applicable.

### Consent for publication

Not applicable.

### Availability of data and materials

The right heart catheterization (RHC) dataset is publicly available at https://hbiostat.org/data/, courtesy of the Vanderbilt University Department of Biostatistics.

### Competing interests

The authors declare that they have no competing interests.

### Funding

This research received no specific grant from any funding agency in the public, commercial, or not-for-profit sectors.

### Authors' contributions

WH performed all the analyses and wrote the manuscript. SA, JT, and TS supervised the study and critically revised the entire manuscript. The authors read and approved the final manuscript.

### Acknowledgments

We would like to thank Editage [http://www.editage.com] for editing and reviewing this manuscript for English language.

**Figure legends**

**Fig. 1** Box plots of 1000 estimates for the average treatment effect (ATE) with continuous covariates.

**Fig. 2** Box plots of 1000 estimates for the average treatment effect (ATE) with binary covariates.